\documentclass[aps,pre,amsmath,amssymb,twocolumn]{revtex4}
\usepackage{graphicx}
\usepackage{amssymb}
\usepackage{amsmath}
\usepackage{epstopdf}
\newcommand{\ba}{\begin{array}{l}}
\newcommand{\ea}{\end{array}}
\newcommand{\re}[1]{(\ref{#1})}
\newcommand{\ci}[1]{\cite{#1}}
\newcommand{\be}{\begin{equation}}
\newcommand{\ee}{\end{equation}}
\newcommand{\lab}[1]{\label{#1}}

\begin{document}
\title{\textbf{Soliton generation in PT-symmetric optical fiber networks}}
\author{M.E. Akramov$^b$, K.K. Sabirov$^{c}$, O.V. Karpova$^{a}$, D.U. Matrasulov$^{a}$, S. Usanov$^d$}
\affiliation{ $^a$Turin Polytechnic University in Tashkent, 17
Niyazov Str.,
100095,  Tashkent, Uzbekistan\\
$^b$Physics Department, National University of Uzbekistan, Vuzgorodok, Tashkent
100174,Uzbekistan\\
$^c$Tashkent University of Information Technology, Amir Temur Avenue 108, Tashkent 100200, Uzbekistan\\
$^d$Yeoju Technical Institute in Tashkent, 156 Usman Nasyr Str., 100121, Tashkent, Uzbekistan}

\begin{abstract}
 We consider the problem of soliton generation in PT-symmetric optical fiber networks, where soliton dynamics is governed by  nonlocal nonlinear Schrodinger equation on metric graphs. Exact formulae for the number of generated solitons are derived for the cases, when the problem is integrable. Numerical solutions are obtained for the case, when integrability is broken.
\end{abstract}

\maketitle

\section{Introduction}
The problem of soliton generation in optical fibers is of fundamental and practical importance for modern optoelectronics and information technologies. It were Hasegawa and Tappert \ci{Hasegawa}, who proposed first to use
optical solitons as carriers of information in high-speed communication
systems in early seventies of the last century. Further development of the idea  later led to advanced optoelectronic and information technologies based on the use of solitons in optical fibers (see, e.g., Refs.  \ci{Fiber1}-\ci{Kivsharbook} for review).
Dynamics of generated solitons strongly depends on the shape of the initial pulse profile. This makes choosing initial pulse profile effective tool for tuning the soliton propagation.  Mathematically, the problem of soliton generation is reduced to the
Cauchy problem for the nonlinear evolution equation, governing the dynamics of soliton. An important task arising in this context, besides soliton dynamics, is finding the  number of
generated solitons using given initial condition. In case of long(unbranched) fibers such problem was studied in the
Refs.\ci{Burzlaff}-\ci{Zhong}. In \ci{Burzlaff}, where an effective
method for computing the number of  generated solitons is proposed.
Extension of the approach for  other initial pulse profiles
was proposed later in \ci{Kivshar}.  Mathematical treatment of soliton
generation on a half line  was
considered  \ci{Fokas}.  Generation in optical solitons in
fibers with a dual-frequency input was considered in \ci{Panoiu}. Soliton generation and their instability are investigated in a system of two
parallel-coupled fibers, with a pumped (active) nonlinear
dispersive core and a lossy (passive) linear one in \ci{Malomed1}. In this paper we address the problem of generation PT-symmetric solitons described in terms of nonlocal nonlinear Schrodinger (NNLS) equation . The latter has attracted much attention since from pioneering paper by Ablowitz and Muslimani \cite{AM2013}, where soliton solution NNLS equation was obtained and integrability of the problem was shown. Different aspects of NNLS equation have been studied since from that (see, Refs.  \ci{AM2013}-\ci{Yang} for review of the progress in the topic). Recently, nonlocal PT-symmetric solitons were studied by modelling them in terms of NNLS equation on metric graphs \cite{Mashrab2021}. Physically, solitons described in terms of NNLS equation can be realized in optical materials providing self-induced gain and loss. In this paper we address the problem of generation PT-symmetric nonlocal solitons in a network of optical fibers. Such PT-symmetric structure  can be constructed in the form of  optical waveguide network where each branch has "gain-and-loss". Motivation for the study of soliton generation in PT-symmetric optical waveguide networks comes from the fact that by choosing initial pulse configuration and network topology properly, one can tune the generation and propagation processes. This paper is organized as follows. In the next section we present formulation and solution of the problem for linear, i.e., unbranched fiber, described in terms of NNLS equation on a line. In section III we briefly recall NNLS equation on networks. Section IV presents extension of the soliton generation problem to the case of networks by considering star and tree-branched networks. Finally, the section V provides some concluding remarks.

\section{Soliton generation in linear PT-symmetric optical fibers}

A nonlocal nonlinear Schrodinger
equation providing PT-symmetry was proposed first by Ablowitz and
Muslimani in \ci{AM2013} and attracted much attention in different contexts. It can be written as
\begin{equation}
i\frac{\partial}{\partial t}q(x,t)=\frac{\partial^2}{\partial
x^2}q(x,t)+2 q^2(x,t)q^*(-x,t).\label{nlse0}
\end{equation}
Introducing  the PT-symmetric self-induced
potential, $V=-2q(x,t)q^*(-x,t)$, one can write Eq.\re{nlse0} in form of the following linear Schrodinger equation:
\begin{equation}
\frac{\partial}{\partial t}q(x,t)=-i\frac{\partial^2}{\partial
x^2}q(x,t)+iV(x,t)q(x,t),\label{nlse01}
\end{equation}
 Due to the PT-symmetry of potential $V(x,t)$, given by the relation $V(x,t) = V^*(-x,t)$, Eq.\re{nlse01} can be considered as the PT-symmetric Schrodinger equation. We note that from the physical viewpoint, Eq.\re{nlse01} describes the PT-symmetric optical solitons propagating in optical waveguide having "gain-and-loss" structure.
A single-soliton solution of Eq.\re{nlse0} was derived in  \ci{AM2013} and can be written as
\be
q(x,t)=-\frac{2(\eta_1+\bar{\eta}_1)e^{i\bar{\theta}_1}e^{-4i\bar{\eta}^2_1t}e^{-2\bar{\eta}_1x}}{1+e^{i(\theta_1+\bar{\theta}_1)} e^{4i(\eta^2_1-\bar{\eta}^2_1)t} e^{-2(\eta_1+\bar{\eta}_1)x}}.
\lab{sol01}
\ee
An important task of soliton generation problem is finding the number of solitons generated for a given initial pulse profile. From the mathematical viewpoint, such a task represents initial value (Cauchy) problem for a given initial pulse profile.   An effective method for solving such task was proposed in \ci{Burzlaff}, which was later applied for different types of pulse profile in \cite{Kivshar,Panoiu}.  Starting point in calculation the number of solitons generated for a given initial pulse profile is the Zakharov-Shabat problem. For Eq.\re{nlse0} Zakharov-Shabat problem is given in terms of the following AKNS system:
\begin{eqnarray}
\frac{\partial v^{(1)}}{\partial x}=-ik
v^{(1)}+ q(x,0)v^{(2)},\nonumber\\
\frac{\partial v^{(2)}}{\partial x}=ik
v^{(2)} - q^*(-x,0)v^{(1)},\label{eq1}
\end{eqnarray}
where $q(x,0)$ is the initial condition (initial pulse
profile) for NNLS equation. Let us consider the special
family of the initial potentials
\begin{gather}
q(x,0)=Q(x,0)e^{i(\delta+\pi/2)},\nonumber \\
q^*(-x,0)=Q(-x,0)e^{-i(\delta+\pi/2)},\label{eq2}
\end{gather}
\begin{figure}[t!]
\includegraphics[scale=0.3]{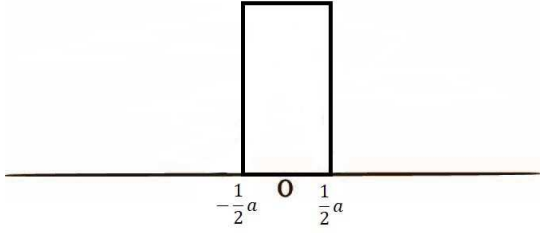}
\caption{Initial pulse profile}
\end{figure}
where $Q(x,0)$ is the real function and
$\delta\,(0\leq\delta\leq2\pi)$ is arbitrary constant. One
can show that  the transformations
\begin{equation}
v^{(1)}\to V^{(1)}e^{i\gamma},\quad
v^{(2)}\to
V^{(2)}e^{i(\gamma-\delta)}\label{eq3}
\end{equation}
lead to the following eigenvalue problem
\begin{eqnarray}
\frac{\partial V^{(1)}}{\partial x}=-ik
V^{(1)}+iQ(x,0)V^{(2)},\nonumber\\
\frac{\partial V^{(2)}}{\partial x}=ik
V^{(2)} + iQ(-x,0)V^{(1)}.\label{eq4}
\end{eqnarray}
Following the Ref. \cite{kivshar}, one can  define the
number of the zeros of the Jost coefficients $a(k)$ at
$k=0$.

If the initial condition is symmetric to the point $x = 0$: $Q(x,0)=Q(-x,0)$ then
the formal solution of Eq.\eqref{eq4} with $k=0$ are
\begin{widetext}
\begin{gather}
V^{(1)}(x,0)=\exp\left(-iS(x)\right)\left(C^{(1)}\underset{-\infty}{\overset{x}{\int}}Q(x',0)\exp\left(2iS(x')\right)dx'+C^{(2)}\right),\nonumber\\
V^{(2)}(x,0)=-iC^{(1)}\exp\left(iS(x)\right)-V^{(1)},\label{eq5}
\end{gather}
\end{widetext}
where
$$ S(x)=\underset{-\infty}{\overset{x}{\int}}Q(x',0)dx',$$
If one chooses $V^{(1)}(x,0)\to0$ for $x\to-\infty$ and
$V^{(2)}(x,0)\to 0$ for $x\to+\infty$,  then $C^{(2)}=0,$ and we have
\begin{widetext}
\begin{eqnarray}
&a(0)=\underset{x\to+\infty}{\lim}V^{(2)}(x,0)=\nonumber\\
&=-iC^{(1)}\left(\exp(iS_{0})-i\exp(-iS_{0})\underset{-\infty}{\overset{+\infty}{\int}}Q(x,0)\exp(2iS(x))dx\right)=-iC^{(1)}\cos
S_{0},\label{eq8}
\end{eqnarray}
\end{widetext}
where
\begin{equation}
S_{0}=\underset{-\infty}{\overset{+\infty}{\int}}Q(x,0)dx.\label{eq10}
\end{equation}
From Eqs. (\ref{eq8}) for the soliton number we
get
\begin{equation}
N=\langle\frac{1}{2}+\frac{S_{0}}{\pi}\rangle.\label{eq11}
\end{equation}
Noting that for the initial  pulses given by Eq. \eqref{eq2}  for
any $x$ and with $Q(x,0)>0$
\begin{equation}
S_{0}\equiv\underset{-\infty}{\overset{+\infty}{\int}}Q(x,0)dx=\underset{-\infty}{\overset{+\infty}{\int}}|q(x,0)|dx=F.\label{eq12}
\end{equation}
we have from Eqs.\eqref{eq11} and \eqref{eq12}
\begin{equation}
N=\langle\frac{1}{2}+\frac{F}{\pi}\rangle.\label{eq13}
\end{equation}


Here we consider number of generated solitons for the rectangular initial pulse profile:
\begin{eqnarray}
q(x,0)=
\begin{cases}
0, & \text{for}\;\;\;  |x|>\frac{1}{2}a \\
b, & \text{for}\;\;\;  |x|\leq\frac{1}{2}a
\end{cases}\quad\quad\quad b>0.\nonumber
\end{eqnarray}

Using the above approach for this profile leads to
\begin{equation}
F=\int_{-\infty}^{+\infty}|q(x,0)|dx=ab, \quad N=\left<\frac{1}{2}+\frac{ab}{\pi} \right>.\nonumber
\end{equation}
This equation provides relation between the initial pulse profile and number of generated solitons, described by the PT-symmetric nonlocal nonlinear Schrodinger equation \re{nlse0}.

\section{Soliton generation in star-shaped optical waveguide network}
\begin{figure}[t!]
\includegraphics[scale=0.3]{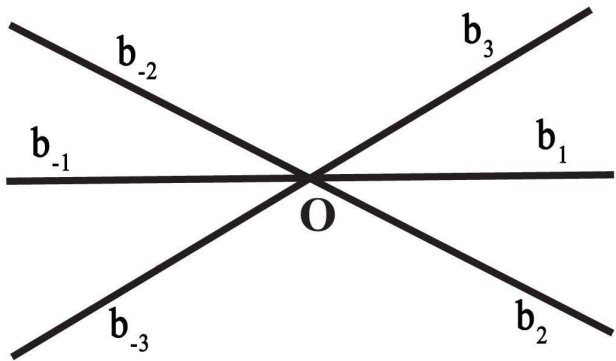}
\caption{Star graph with six bonds}
\end{figure}
The above approach can be applied for soliton generation in branched waveguides, by modelling these lattes in terms of so-called metric graphs, which are the systems of wires connected to each other at the nodes (vertices) according to some rule, called topology of a graph. Such process is described in terms of the nonlocal nonlinear Schrodinger equation on graphs studied recently in \cite{Mashrab2021}.  We note that evolution equations on metric graphs attracted much attention during the last decade (see, Refs.\cite{Zarif} -\cite{SGN2020}). Such NNLS equation on a six-bond, star branched graph (see, Fig. 2) can be written as
\begin{equation}
i\frac{\partial}{\partial t} q_{\pm j}(x,t)=\frac{\partial^2}{\partial x^2} q_{\pm j}(x,t) + \sqrt{\beta_j \beta_{-j}} q^2_{\pm j}(x,t)q_{\mp j}^*(-x,t),
\lab{nnlse1}
\end{equation}
where $q_{\pm j}(x,t)$ at $x\in b_{\pm j}$ and $j=1,2,3$. Eq.\re{nnlse1} is written  on the each bond of
the star graph with six bonds $b_{\pm j}$ (see, Fig. 2), for which a coordinate $x_{\pm j}$ is assigned. The origin of coordinates is chosen at the vertex, for bond $b_{-j}$ we put
$x_{-j}\in (-\infty,0]$ and for $b_j$ we fix $x_j\in[0,+\infty)$.
An important feature of Eq.\re{nnlse1} comes from the fact that it is a system of coupled nonlocal nonlinear Schrodinger equations in which components of $q_{\pm j}$ are mixed in nonlinear term. In usual NLSE on graphs, such mixing does not appear explicitly, but caused by the vertex boundary conditions. Complete task formulation for NNLS equation on metric star graph requires imposing the boundary conditions at the node (vertex). Such boundary conditions can be derived, e.g., from physically relevant conservation laws. A set of the vertex boundary conditions following from the norm and energy conservation can be written as \cite{Mashrab2021}:
\begin{widetext}
\begin{gather}
\alpha_{1} q_{1}(x,t)|_{x=0}=\alpha_{-1}q_{-1}(x,t)|_{x=0}=\alpha_{2}q_{2}(x,t)|_{x=0}=\alpha_{-2} q_{-2}(x,t)|_{x=0}=\alpha_{3} q_{3}(x,t)|_{x=0}=\alpha_{-3} q_{-3}(x,t)|_{x=0},\nonumber\\
\left.\frac{1}{\alpha_{1}}\frac{\partial}{\partial
x}q_{1}(x,t)\right|_{x=0}+
\left.\frac{1}{\alpha_{2}}\frac{\partial}{\partial
x}q_{2}(x,t)\right|_{x=0}+
\left.\frac{1}{\alpha_{3}}\frac{\partial}{\partial
x}q_{3}(x,t)\right|_{x=0}=\nonumber\\
\left.\frac{1}{\alpha_{-1}}\frac{\partial}{\partial
x}q_{-1}(x,t)\right|_{x=0}+
\left.\frac{1}{\alpha_{-2}}\frac{\partial}{\partial
x}q_{-2}(x,t)\right|_{x=0}+
\left.\frac{1}{\alpha_{-3}}\frac{\partial}{\partial
x}q_{-3}(x,t)\right|_{x=0}
.\label{bc1}
\end{gather}
\end{widetext}
The problem given by Eqs.\re{nnlse1}-\re{bc1} were recently studied in detail in the Ref.\cite{Mashrab2021}, where constraints providing the integrability of the NNLS equation on graphs have been derived in terms of the nonlinearity coefficients, $\beta_{\pm j}$. Here we briefly recall these results, which will be utilized for solving of soliton generation problem. Below, using the approach applied in the previous section, we demonstrate derivation of expression for the number of solitons generated in a metric star graph.
Let $q(x,t)$ is the solution of Eq.\re{nlse0} and the following constraints are fulfilled:
\begin{gather}
\frac{\alpha_{\pm j}}{\alpha_{1}}=\sqrt{\frac{\beta_{\pm j}}{\beta_{1}}},\nonumber\\
\frac{1}{\beta_{1}}+\frac{1}{\beta_{2}}+\frac{1}{\beta_{3}}=
\frac{1}{\beta_{-1}}+\frac{1}{\beta_{-2}}+\frac{1}{\beta_{-3}}.\label{constrain1}
\end{gather}
Then solution of  NNLS equation (\ref{nnlse1}),  on metric star graph fulfilling the boundary conditions \eqref{bc1} can be written as
$$q_{\pm j}(x,t)=\sqrt{\frac{2}{\beta_{\pm j}}}q(x,t)$$

\begin{figure}[t!]
\label{star graph}
\includegraphics[scale=0.3]{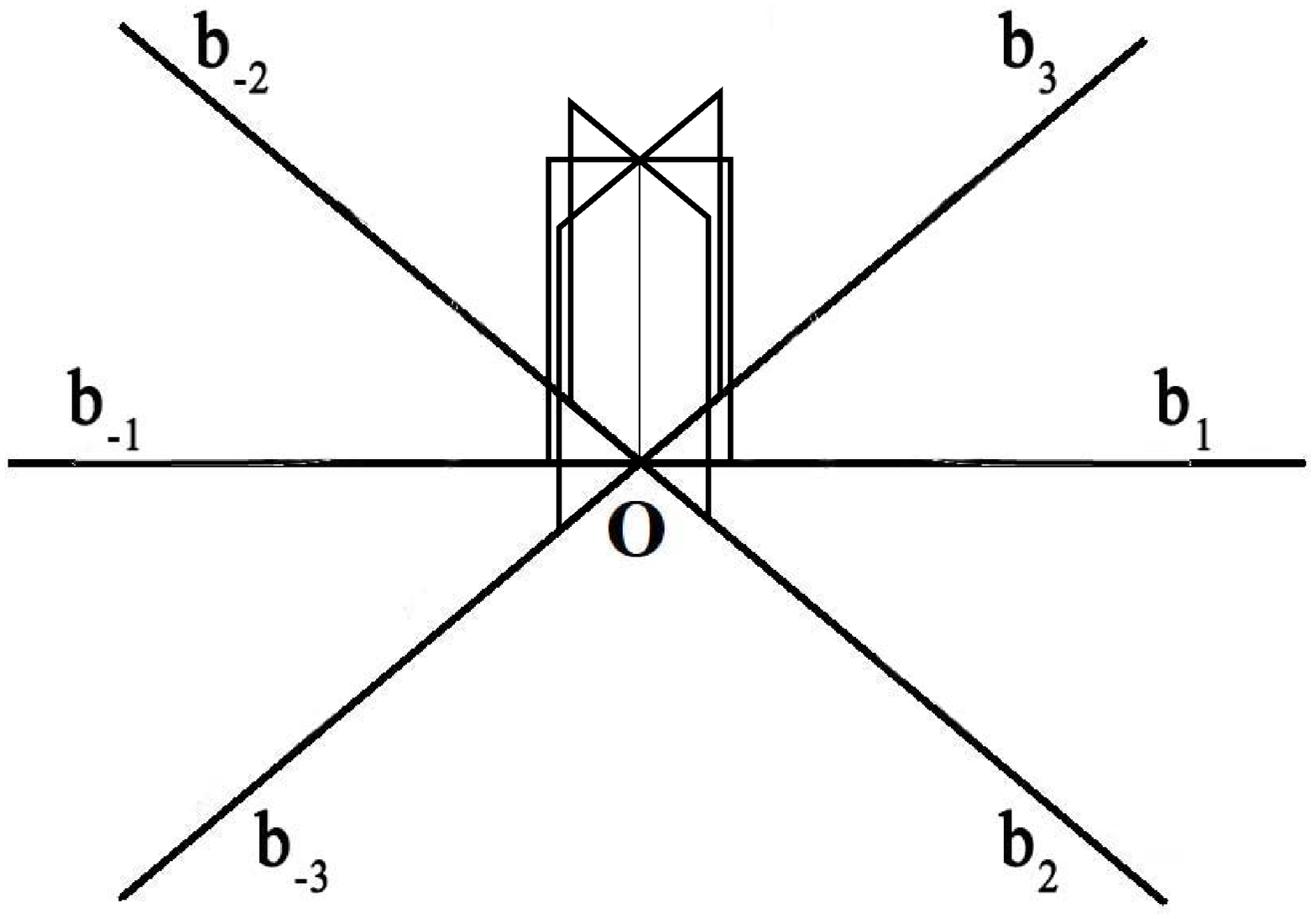}
\caption{Initial pulse profile for the star graph}
\end{figure}
For the soliton  solution given by Eq. (\ref{sol01}), the solution of NNLS equation a star graph can be written as
\begin{eqnarray}
q_{\pm j}(x,t)=-\sqrt{\frac{2}{\beta_{\pm j}}}\frac{4\eta
e^{i\bar{\varphi}}e^{-4i\eta^2t}e^{-2\eta
x}}{1+e^{i(\varphi+\bar{\varphi})}e^{-4\eta x}}.\label{sol1}
\end{eqnarray}
$\varphi,\,\bar{\varphi},\,\eta$ are arbitrary complex constants.

Here we will provide brief derivation of the relation
between the number of generated solitons and the initial pulse
profile in a branched optical waveguide, which is modeled in terms of the star graph presented in Fig.3. Consider the following Zakharov-Shabat problem for NNLS equation,\re{nnlse1}:
\begin{gather}
\frac{\partial v_{\pm j}^{(1)}}{\partial x}=-ik
v_{\pm j}^{(1)}+ \sqrt{\frac{\beta_{\pm j}}{2}} q_{\pm j}(x,0)v_{\pm j}^{(2)},\nonumber\\
\frac{\partial v_{\pm j}^{(2)}}{\partial x}=ik
v_{\pm j}^{(2)} - \sqrt{\frac{\beta_{\mp j}}{2}}q_{\mp j}^*(-x,0)v_{\pm j}^{(1)},\label{eq1b}
\end{gather}
where $q_{\pm j}(x,0)$ are the initial conditions (initial pulse
profiles) for Eq.\re{nnlse1}. Introducing the special
family of the initial potentials given by
\begin{gather}
q_{\pm j}(x,0)=Q_{\pm j}(x,0)e^{i(\delta_{\pm j}+\pi/2)},\nonumber \\
q^*_{\pm j}(-x,0)=Q_{\pm j}(-x,0)e^{-i(\delta_{\pm j}+\pi/2)}\label{eq2b}
\end{gather}
where $Q_{\pm j}(x,0)$ are the real functions and
$\delta_{\pm j}\,(0\leq\delta_{\pm j}\leq2\pi)$ are arbitrary constants, one can show that  the transformations
\begin{equation}
v_{\pm j}^{(1)}\to V_{\pm j}^{(1)}e^{i\gamma_{\pm j}},\quad
v_{\pm j}^{(2)}\to
V_{\pm j}^{(2)}e^{i(\gamma_{\pm j}-\delta_{\pm j})}\label{eq3b}
\end{equation}
lead to the following eigenvalue problem
\begin{gather}
\frac{\partial V_{\pm j}^{(1)}}{\partial x}=-ik
V_{\pm j}^{(1)}+i\sqrt{\frac{\beta_{\pm j}}{2}}Q_{\pm j}(x,0)V_{\pm j}^{(2)},\nonumber\\
\frac{\partial V_{\pm j}^{(2)}}{\partial x}=ik
V_{\pm j}^{(2)} + i\sqrt{\frac{\beta_{\mp j}}{2}}Q_{\mp j}(-x,0)V_{\pm j}^{(1)}.\label{eq4b}
\end{gather}
From a physical viewpoint, the generation of the single quiescent
soliton will occur with a smaller energy than the soliton pair.
Therefore, following the Ref. \cite{kivshar}, we will define the
number of the zeros of the Jost coefficients $a_{\pm j}(k)$ at
$k=0$. If the initial condition is symmetric with respect the point $x = 0$:
$Q_{\mp j}(-x,0)=\sqrt{\frac{\beta_{\pm j}}{\beta_{\mp j}}}Q_{\pm j}(x,0)$.

The formal solutions of Eq. (\ref{eq4b}) with $k=0$ are
\begin{widetext}
\begin{gather}
V_{-j}^{(1)}(x,0)=\exp\left(-iS_{-j}(x)\right)\left(C_{-j}^{(1)}\underset{-\infty}{\overset{x}{\int}}Q_{-j}(x',0)\exp\left(2iS_{-j}(x')\right)dx'+C_{-j}^{(2)}\right),\nonumber\\
V_{-j}^{(2)}(x,0)=-iC_{-j}^{(1)}\exp\left(iS_{-j}(x)\right)-V_{-j}^{(1)},\nonumber \label{eq5b}\\
V_{j}^{(1)}(x,0)=\exp\left(-iS_{j}(x)\right)\left(C_{j}^{(1)}\underset{0}{\overset{x}{\int}}Q_{j}(x',0)\exp\left(2iS_{j}(x')\right)dx'+C_{j}^{(2)}\right),\nonumber\\
V_{j}^{(2)}(x,0)=-iC_{j}^{(1)}\exp\left(iS_{j}(x)\right)-V_{j}^{(1)},\nonumber
\end{gather}
\end{widetext}
$$ S_{-j}(x)=\sqrt{\frac{\beta_{-j}}{2}}\underset{-\infty}{\overset{x}{\int}}Q_{-j}(x',0)dx'$$
and
$$
S_{j}(x)=\sqrt{\frac{\beta_{j}}{2}}\underset{0}{\overset{x}{\int}}Q_{j}(x',0)dx'.\label{eq7b}
$$
If one chooses $V_{-j}^{(1)}(x,0)\to 0$ for $x\to-\infty$ and
$V_{j}(x,0)\to 0$ for $x\to+0$,  then $C_{\pm j}^{(2)}=0,$ and we have
\begin{widetext}
\begin{eqnarray}
&a_{-j}(0)=\underset{x\to-0}{\lim}V_{-j}^{(2)}(x,0)=\nonumber\\
&=-iC_{-j}^{(1)}\left(\exp(iF_{-j})-i\exp(-iF_{-j})\underset{-\infty}{\overset{0}{\int}}Q_{-j}(x,0)\exp(2iS_{-j}(x))dx\right)=-iC_{-j}^{(1)}\cos
F_{-j},\label{eq8b}\\
&a_{j}(0)=\underset{x\to+\infty}{\lim}V_{j}^{(2)}(x,0)=\nonumber\\
&=-iC_{j}^{(1)}\left(\exp(iF_j)-i\exp(-iF_j)\underset{0}{\overset{+\infty}{\int}}Q_{j}(x,0)\exp(2iS_{j}(x))dx\right)=-iC_{j}^{(1)}\cos
F_{j}.\label{eq9b}
\end{eqnarray}
\end{widetext}
Noting that for the initial  pulses given by Eq.\eqref{eq2b}  for
any $x$ and with $Q_{\pm j}(x,0)>0$
\begin{equation}
F_{\pm j}=\sqrt{\frac{\beta_{\pm j}}{2}}\underset{b_{\pm j}}{\int}Q_{\pm j}(x,0)dx=\sqrt{\frac{\beta_{\pm j}}{2}}\underset{b_{\pm j}}{\int}|q_{\pm j}(x,0)|dx.\label{eq10b}
\end{equation}
From Eqs. (\ref{eq8b}) and (\ref{eq9b}) for the soliton number we
get
\begin{equation}
N=\bigg\langle 3+\frac{\sum_{j=1}^3 (F_{-j}+F_{j})}{\pi}\bigg\rangle.\label{NN}
\end{equation}

Now consider the star graph with rectangle initial pulse (see, Fig 3). For such profile, the initial condition is given at the vertex and can be written as $q_{\pm j}(x,0)=\sqrt{\frac{2}{\beta_{\pm j}}}\psi_{\pm j}(x)$:
\begin{gather}
\psi_{-j}(x)=
\begin{cases}
0, & \text{for}\;\;\;  x<-\frac{1}{2}a \\
b, & \text{for}\;\;\;  -\frac{1}{2}a\leq x\leq 0
\end{cases}, \nonumber\\
\psi_{j}(x)=
\begin{cases}
0, & \text{for}\;\;\;  x>\frac{1}{2}a \\
b, & \text{for}\;\;\;  0\leq x\leq\frac{1}{2}a
\end{cases},\nonumber
\end{gather}
where $b>0$.

The number of generated solitons
\begin{gather}
F=\sum_{j=1}^3 (F_{-j}+F_{j})=\sum_{j=1}^3 \bigg(\sqrt{\frac{\beta_{-j}}{2}} \int_{b_{-j}} |q_{-j}(x,0)|dx+\nonumber\\
\sqrt{\frac{\beta_{j}}{2}} \int_{b_j} |q_j(x,0)|dx\bigg)= 3ab,\nonumber\\
N=\left<3+\frac{3ab}{\pi} \right>.\lab{s1}\nonumber
\end{gather}

Another initial pulse profile is the Gaussian one given by
\begin{eqnarray}
q_{\pm j}(x,0)=\sqrt{\frac{2}{\beta_{\pm j}}}A\exp\left[-\frac{1}{2}(1-i\alpha)\left(\frac{x}{\sigma}\right)^{2m}\right].\label{Gaussian}
\end{eqnarray}
Using the above approach for this profile leads to
\begin{gather}
F=\sum_{j=1}^3 \bigg(\sqrt{\frac{\beta_{-j}}{2}} \int_{b_{-j}} |q_{-j}(x,0)|dx+\nonumber\\
\sqrt{\frac{\beta_{j}}{2}} \int_{b_j} |q_j(x,0)|dx\bigg)=\frac{3\cdot 2^{\frac{1}{2m}}A\sigma}{m}\Gamma\left(\frac{1}{2m}\right),
\nonumber\\
N=\left<3+\frac{F}{\pi} \right>.\nonumber
\end{gather}

\begin{figure}[t!]
\label{radiation}
\centering
\includegraphics[scale=0.7]{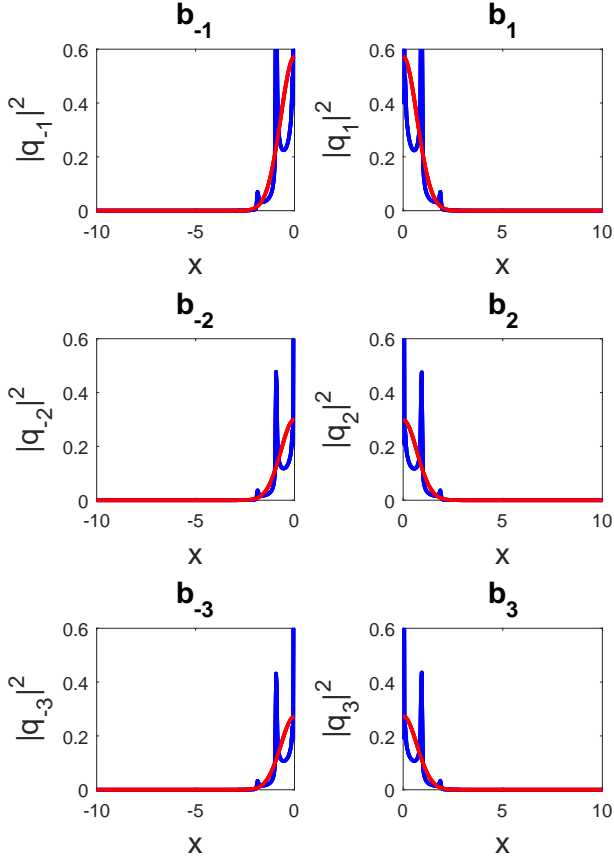}
\caption{Radiation of the pulse profile at $t=0$ (red line) and $t=0.0022$ (blue line) on the star graph.}
\end{figure}
We note that Eq.\eqref{NN} for the number of generated solitons is derived under the assumption that the sum rule in Eqs.\re{constrain1}, which is equivalent to the integrability of NNLS equation on graph (see, the Ref.\cite{Mashrab2021}). For the case, when sum rule is broken, one needs to solve the problem numerically, by imposing the initial conditions given by Eq.\eqref{Gaussian}. The plots of the numerically obtained solution are presented in Fig.4 for the time moments, $t=0$ and $t=0.0022$. Discretization scheme proposed in \ci{AM2014} is used for numerical solution of the initial value problem for NNLS equation on a star graph. An important feature of the soliton generation, i.e., breaking of the initial pulse profile due to the radiation can be observed from the plots of Fig.4.

\section{Extending for the tree graph}

The central branch, i.e. the branch st the middle of the graph is chosen as an origin of coordinates. Then the bonds can be determined as
$b_{-1}, \; b_{-1nm}\sim (-\infty; 0]$,
$b_{-1n}\sim [-L_{1n}; 0]$,
$b_{1n}\sim [0; L_{1n}]$,
$b_{1},\; b_{1nm}\sim [0; +\infty)$, where $L_{1n}$ are the lengths of $b_{\pm 1n}$ bonds and $n=1,2$, $m=1,2$. Here the "+" sign is for right-handed bonds and the "-" sign is for left-handed bonds from the center of the tree graph.
Soliton solutions on each bond can be written as
\begin{eqnarray}
q_{\pm 1}(x,t)=\sqrt{\frac{2}{\beta_{\pm 1}}}q(x+S_{\pm 1},t),\nonumber\\
q_{\pm 1m}(x,t)=\sqrt{\frac{2}{\beta_{\pm 1m}}}q(x+S_{\pm 1m},t),\label{constr01}\\
q_{\pm 1mn}(x,t)=\sqrt{\frac{2}{\beta_{\pm 1mn}}}q(x+S_{\pm 1mn},t).\nonumber
\end{eqnarray}
Note that solutions given by Eq.\re{constr01} hold true provided the following sum rules are fulfilled \cite{Mashrab2021}:
\begin{gather}
\frac{1}{\beta_{\pm 1}}=\frac{1}{\beta_{\mp 11}}+\frac{1}{\beta_{\mp 12}}, \nonumber\\
\frac{1}{\beta_{\pm 1n}}=\frac{1}{\beta_{\pm 1n1}}+\frac{1}{\beta_{\pm 1n2}}.
\end{gather}

Furthermore, we choose the initial pulse profile at each vertex
($q_e(x,0)=\sqrt{\frac{2}{\beta_e}}\psi_e(x)$) in the forms
\begin{gather}
\psi_{\pm 1}(x)=
\begin{cases}
0, & |x|>\frac{1}{2}a \\
A, & 0\leq |x|\leq \frac{1}{2}a
\end{cases},\nonumber\\
\psi_{\pm 1n}(x)=
\begin{cases}
A, & 0\leq |x| \leq \frac{1}{2}a \\
0, & \frac{1}{2}a<|x|<L_{1n}-\frac{1}{2}a\\
A_n, & L_{1n}-\frac{1}{2}a \leq |x| \leq L_{1n}
\end{cases},\nonumber\\
\psi_{\pm 1nm}(x)=
\begin{cases}
A_n, & 0\leq |x| \leq \frac{1}{2}a \\
0, &  |x|>\frac{1}{2}a
\end{cases}.\nonumber
\end{gather}

Then for the number solitons we have
\begin{eqnarray}
N=\left< 7+\frac{\sum_{s=-2}^2 F_s}{\pi} \right>, \nonumber\label{number03}
\end{eqnarray}
where
\begin{gather}
F_0=3aA,\quad F_{\pm 1}=\frac{3aA_1}{2},\quad F_{\pm 2}=\frac{3aA_2}{2}.\nonumber
\end{gather}
\begin{figure}[t!]
\label{tree_graph}
\includegraphics[scale=0.06]{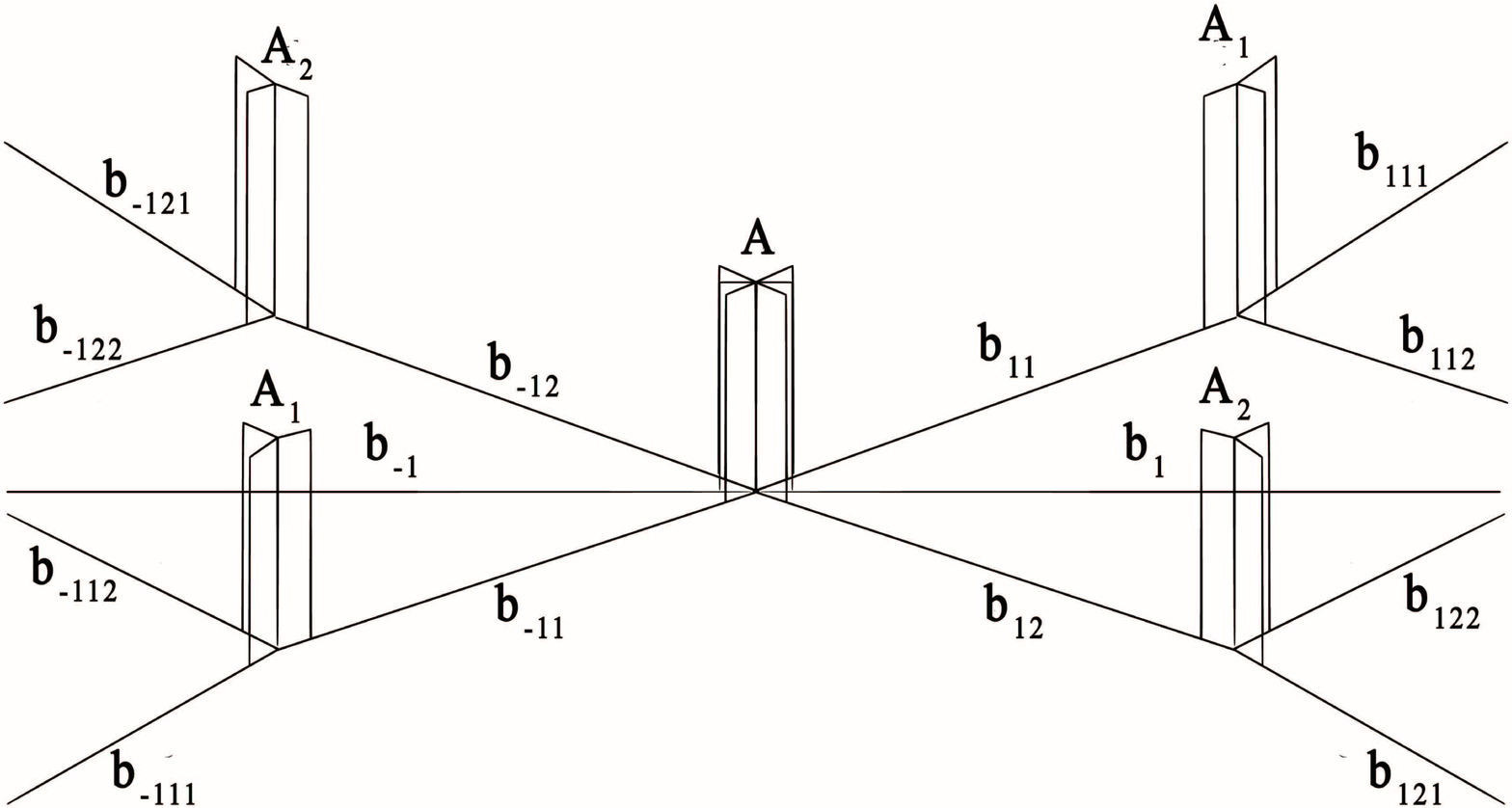}
\caption{Initial pulse profile for the tree graph}
\end{figure}
Again, for the case, when the constraints given by Eq.\re{constr01}, NNLS equation should be solved numerically and Eq.\re{number03} cannot be used for finding the number of solitons generated.
\section{Conclusions}
In this paper we studied the problem of soliton generation for PT-symmetric optical waveguides and their networks described in terms of NNLS equation. The problem of finding the number of generated solitons for a given initial pulse profile is reduced to the Cauchy problem for NNLS equation on a line and on metric graphs, where the initial condition is give in terms of the initial pulse. Exact expression for the number of solitons generated is derived. In case of optical waveguide networks, the problem is solved for star- and tree-branched networks. The results obtained in this paper and proposed models can be applied for the problem of  tunable generation of solitons in branched optical fiber networks providing PT-symmetry via "gain-loss" property.  Experimental realization of such a model is of importance for engineering and practical implementation of PT-symmetric optical fiber networks, capable to generate solitonic pulses and tunable signal propagation. Although the above treatment deals with star- and tree graphs, the method can be extended for arbitrary graph topologies having semi-infinite incoming and outgoing bonds.

\end{document}